\documentclass[twocolumn,showpacs,preprintnumbers,amsmath,amssymb,prb]{revtex4}
\usepackage{graphicx}% Include figure files
\usepackage{dcolumn}% Align table columns on decimal point
\usepackage{bm}% bold math
\usepackage{colortbl}

\begin{document}

\preprint{APS/123-QED}

\title{Quantum breathing mode of trapped bosons and fermions at arbitrary coupling}% Force line breaks with \\
\author{S.~Bauch}
\author{K.~Balzer}
\author {C.~Henning}
\author{M.~Bonitz}
 \email{bonitz@physik.uni-kiel.de}
\affiliation{%
Institut f\"ur Theoretische Physik und Astrophysik\\
Christian-Albrechts-Universit\"at zu Kiel, D-24098 Kiel, Germany
}%

\date{\today}% It is always \today, today,
             %  but any date may be explicitly specified

\begin{abstract}
Interacting particles in a harmonic trap are known to possess a radial collective oscillation -- the breathing mode (BM). We show that a quantum system has {\em two BMs} and analyze their properties by exactly solving the time-dependent Schr\"odinger equation. We report that the frequency of one BM changes with system dimensionality,
the particle spin and the strength of the pair interaction and propose a scheme that gives direct access to key properties of trapped particles, including their many-body effects.

\end{abstract}

% PACS, the Physics and Astronomy
                             % Cla ssification Scheme.
%\keywords{Suggested keywords}%Use showkeys class option if keyword
                              %display desired
\maketitle

%Introductory part
%----------------------------
\section{Introduction}
The dynamics of trapped quantum systems are of growing interest in many fields, 
including correlated electrons in metal clusters \cite{baletto05} or quantum dots \cite{afilinov-etal.prl01,afilinov-etal.pss00,reimann02}
and ultracold Bose and Fermi gases in traps or optical lattices, for recent overviews see e.g. \cite{Bloch2005,Giorgini2008}. Particular attention has recently been devoted to Bose-Einstein condensation in low dimensions \cite{Goerlitz2001} and to the analysis of nonideality (interaction) effects\cite{Menotti2002,Moritz2003,Pedri2008,bloch2008}, including superfluidity and crystallization, see\cite{afilinov-etal.08prb}. 
At the same time, nonideal low-dimensional Bose and Fermi systems present major experimental and theoretical challenges. For a reliable diagnostics of static and time-dependent properties, the collective oscillations of the system play a key role \cite{Menotti2002,Moritz2003,Giorgini2008}. It is the purpose of this paper to show that, among them, the monopole or breathing mode (BM) which is easily excited experimentally \cite{Moritz2003} carries particularly valuable information on the system dimensionality $d$, the spin statistics of the particles and on the form and relative strength $\lambda$ of their pair interaction given by Eqs.~(\ref{eq:v},\ref{lambda}). 

The BM of a system of $N$ particles in a harmonic trap with frequency $\Omega$ is well known in two limiting cases:
in the limit of very strong interparticle repulsion ($\lambda \to \infty $) with an inverse power law potential, $w(|{\bf r}|)\sim r^{-l}$, the particles behave classically being spatially localized, and the BM which describes the radial expansion/contraction of a cloud of well-separated particles is independent of $N$ and $d$  \cite{partoens_etal.jphy97,Henning2008}, whereas its frequency is sensitive to the interaction, $\omega_{\rm BM} = \sqrt{2l+1} \,\Omega$. 
In the second limiting case, that of an ideal quantum gas with $\lambda=0$, the BM corresponds to a periodic expansion/contraction of the wave function with frequency $\omega_{\rm BM}=2 \Omega$, which is, of course, independent of $N$, $d$ and the type of the interaction, see Sec.~\ref{switch_ss}.

This raises the question about $\omega_{\rm BM}$ for {\em arbitrary finite values of the interaction strength} which we answer in this paper. While our results for the BM are representative for any nonideal quantum system in a harmonic trap, here, we concentrate on a complete analysis of the case of Coulomb interaction ($l=1$). We present exact numerical results from solutions of the time-dependent Schr\"odinger equation (TDSE) for two fermions and bosons in $1d$ and $2d$ \cite{l23} and time-dependent Hartree-Fock (TDHF) results for $N=2-4$.
We show that, in fact, there co-exist {\em two independent breathing modes} one of which is $\lambda-$dependent and the relative spectral weight of which varies with $\lambda$. Further, fermions and bosons have the same breathing modes in $1d$, whereas a substantial difference is observed in higher dimensions. 

%------------------------------------------------------------------------------------------------------------------
\section{Model and solution procedure} 
We start by considering two identical particles ($m_1=m_2 \equiv m$, $q_1=q_2 \equiv q$) in a harmonic potential of frequency $\Omega$ with Coulomb repulsion, described by the TDSE of two particles
%--------------------------------------------------------------------------------------------------------------------------------------------------
% equation: two-particle TDSE
%--------------------------------------------------------------------------------------------------------------------------------------------------
\begin {eqnarray}
\bigg[i \frac{\partial}{\partial t} + \frac{1}{2}\left( \frac{\partial^2}{\partial {\bf{r}}^2_1} + \frac{\partial^2}{\partial {\bf{r}}^2_2}\right)  &-& V({\bf{r}}_1, {\bf{r}}_2) \bigg]\Psi({\bf{r}}_1,{\bf{r}}_2,t)=0 \;,\quad
\label{eq:twopart_tdse}
\\
V &=& \frac{{\bf{r}}_1^2}{2} + \frac{{\bf{r}}_2^2}{2} + \frac{\lambda}{|{\bf{r}}_1-{\bf{r}}_2|},
\label{eq:v}
\end {eqnarray}
%--------------------------------------------------------------------------------------------------------------------------------------------------
where the total potential $V$ is the sum of harmonic confinement and Coulomb repulsion.
 The coupling parameter -- the ratio of the mean interaction and single-particle energy -- is given by 
\begin{equation}
\lambda = \frac{q^2 }{4\pi \epsilon_0 l_0}\frac{1}{\hbar \Omega} , 
\label{lambda}
\end{equation}
with the oscillator length $l_0= \sqrt{\hbar/m \Omega}$. Throughout this work, lengths, times and energies will be given in units of $l_0, \Omega^{-1}$ and $\hbar \Omega$, respectively. 
\subsection{Initial conditions}
%-----------------------------------
%initial conditions
%-----------------------------------
Eq.~(\ref{eq:twopart_tdse}) has to be supplemented by an initial condition for the two-particle wave function. 
Here we use either a symmetric ($S$) or anti-symmetric ($A$) function, 
$\Psi^{S,A}_0=\Psi^{S,A}({\bf{r}}_1,{\bf{r}}_2,0)=\pm \Psi^{S,A}({\bf{r}}_2,{\bf{r}}_1,0)$, the symmetry of which is preserved during the time evolution since the hamiltonian is spin-independent. For the actual form of $\Psi^{S,A}_0$ we choose a stationary solution of Eq.~(\ref{eq:twopart_tdse}) which is computed by imaginary time stepping. 
%-----------------------------------
%excitation
%-----------------------------------
To single out time-dependent solutions of Eq.~(\ref{eq:twopart_tdse}) of pure BM-type we use two different excitation methods: (I) a fast switch of the trap frequency $\Omega$ and (II), the response to a periodic modulation of $\Omega$\cite{dipole_comment}. While the former has computational advantages, the latter is more easily realized experimentally \cite{Moritz2003}.
%-----------------------------------
%Num solution
%-----------------------------------
\subsection{Numerical solution of the TDSE}\label{numerics_ss}
We solve the two-particle TDSE~\eqref{eq:twopart_tdse} numerically using two independent methods: i) a standard grid-based Crank-Nicolson scheme (cn) with at least $1000$ grid points in each direction and ii) by expanding the wave function into a basis of oscillator eigenfunctions (OB) using up to $N_b=625$ basis functions. The numerical parameters are adjusted for each value of $\lambda$, such that the results are fully converged with respect to the time step size, the simulation box size and the number of grid points/basis functions. This has to be undertaken very carefully especially in the case of large $\lambda$, where the repulsive Coulomb interaction attains large values and the accurate numerical treatment becomes challenging.

%%----------------------------------------------------------------------------------------------------------------
%% FIRST FIGURE: timeseries, two frequencies
\begin {figure}
\includegraphics[width=0.49\textwidth]{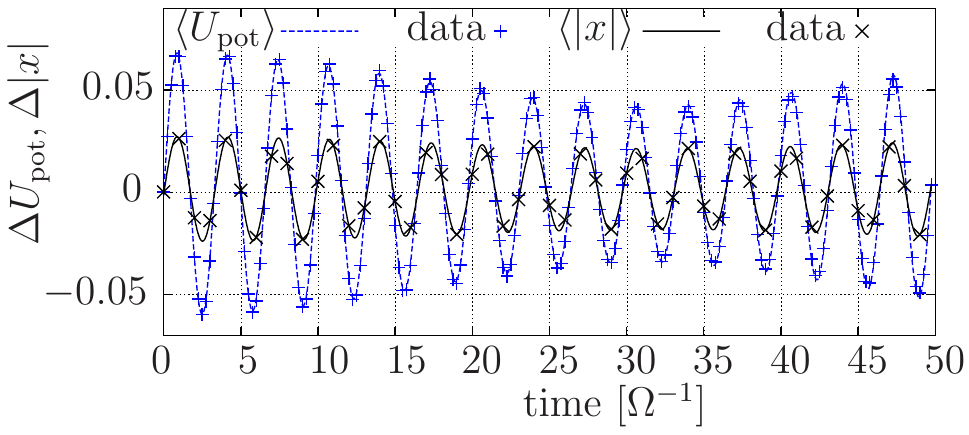}
\caption{(color online) Time evolution of the one-particle potential energy $\Delta U_{\textup{pot}} = \langle U_{\textup{pot}} \rangle(t) - U_{\textup{pot},t=0}$ and $\Delta |x|=\langle |x|\rangle(t) - |x_{t=0}|$ for $N=2$ particles in a $1D$ trap at $\lambda = 1.0$, obtained by solving Eq.~\eqref{eq:twopart_tdse}, symbols. Lines: fit, Eq.~\eqref{eq:fit_2frequencies}, with frequencies $\omega_{r} = 1.901$ and $\omega_{R}=2.0$.}
\label{fig:timeseries}
\end {figure}
%%----------------------------------------------------------------------------------------------------------------

\section{Results}
\subsection{Dynamics following a switch of the confinement}\label{switch_ss} 
Let us start with an antisymmetric initial state $\Psi^{A}_0$ and apply method (I). This is realized by turning off the trap for a short time, typically $\Delta t=0.1\;\Omega^{-1}$, after which it is restored (the explicit time-dependence is not relevant as long as the excitation is spectrally much broader than $\Omega$). During the ``off'' cycle, the Coulomb repulsion drives the particles out of their initial equilibrium state and initiates the BM. 
 
To quantify the oscillatory motion we compute the time-dependent expectation value of the single-particle potential energy $\langle U_{\textup{pot}} \rangle(t)$ with respect to the external trap and, independently, 
 the expectation value $\langle | x|\rangle(t)$ which is directly associated with the monopole oscillation.
Our numerical results confirm that both quantities exhibit identical time dependencies. 
In Fig.~\ref{fig:timeseries} we show the result for $N=2$ in a $1D$ trap at an intermediate coupling, $\lambda=1$, where we expect a quantum BM (QBM) with frequency inbetween the ideal quantum and classical limits, $2$ and $\sqrt{3}$, respectively. However, the simulations reveal a different behavior with evidence of a beating of {\em two oscillations}. This is readily confirmed  by applying a two-frequency fit  
%-----------------------------------------------------
% equation: fit formula for two breathing frequencies
%-----------------------------------------------------
\begin {equation}
f(\omega_r,\omega_R,t)=a\cdot\sin[\omega_r(t-t_0)] + b\cdot \sin[\omega_R (t-t_0')] + f_0,
\label{eq:fit_2frequencies} 
\end {equation}
%------------------------------------------------------------------------------------------------------------------
to our data ($f$ stands for $\langle U_{\textup{pot}} \rangle$ or $\langle | x|\rangle$), where $t_0$ and $t_0'$ indicate phase shifts, $a$ and $b$ are the amplitudes and $\omega_r$ and $\omega_R$ the frequencies of 
the two QBMs, and $f_0$ is the unperturbed (equilibrium) value of  $f$. 
With the values $\omega_{R}=2$ and $\omega_r=1.901$ perfect agreement with the simulations is achieved, cf. lines in Fig.~\ref{fig:timeseries}. 
%-----------------------------------------------------
% equation: Faktorisierung
%-----------------------------------------------------

\begin {figure}
\includegraphics[width=0.49\textwidth]{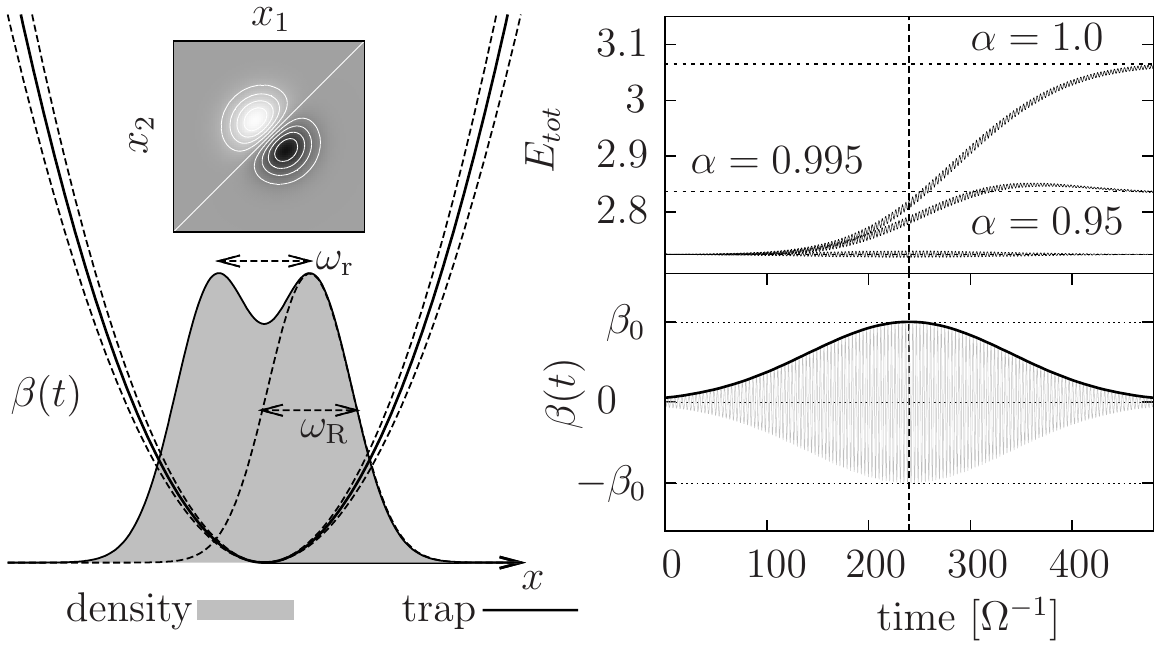}
\caption{Resonance excitation (II) for $\lambda=1.0$ and three frequencies, $\omega_{\textup{ext}}=\alpha \omega_r$ with $\alpha=0.95, 0.995, 1.0$. Left: anti-symmetric initial state (top) and particle density (grey area). Arrows indicate the two breathing motions. Right: time evolution of the total energy for the three external frequencies (top) and the exciting pulse (bottom).}
\label{fig:resonance}
\end {figure}

The origin of the two QBMs becomes immediately clear from the structure of 
Eq.~\eqref{eq:twopart_tdse}. It can be solved with a product ansatz, 
$\Psi(\mathbf{R},\mathbf{r},t) =  \phi(\mathbf{R},t) \cdot  \varphi(\mathbf{r},t)$,
factorizing into functions of the center of mass (CoM) and relative coordinates, ${\bf{R}}=({\bf r}_1 + {\bf r}_2)/2$, and ${\bf r} = {\bf r}_1 - {\bf r}_2$,
resulting in two independent TDSEs:
\begin {equation}
 i \partial_t \varphi (\mathbf{r},t) = \left ( - \frac {\partial^2}{\partial\mathbf{r}^2} + \frac {1}{4} \mathbf{r}^2 + \frac {\lambda}{r} \right ) \varphi (\mathbf{r},t) \;,
\label{eq:relative_tdse}
\end {equation}
whereas $\phi({\bf{R}},t)$ obeys a simple harmonic oscillator problem which is independent of $\lambda$:
\begin {equation}
 i  \partial_t \phi(\mathbf{R},t)  =  \left( -\frac {1}{4} \frac{\partial^2}{\partial\mathbf{R}^2} + \mathbf{R}^2 \right) \phi(\mathbf{R},t) \; .
\label{eq:com_tdse}
\end {equation}
This well-known splitting indicates the existence of two independent motions related to the relative and the CoM problem corresponding, in case of excitation (I), to two QBMs with the characteristic frequencies $\omega_{r}$ and $\omega_{R}$, respectively. Obviously, Eq.~\eqref{eq:com_tdse} leads, in our case, to the (ideal) QBM with frequency $\omega_{R}=2$ because it is just the TDSE for a single harmonic oscillator with mass $m=2$, see \cite{excitation}. Thus, for all couplings $\lambda$, the system (\ref{eq:twopart_tdse}) possesses two QBMs, one with a universal frequency $\omega_{R}=2$ and one with a $\lambda-$dependent frequency $\omega_{r}$; the two modes are depicted by the arrows in the left part of Fig.~\ref{fig:resonance}, for an illustration, see Ref.~\cite{video}. In the ideal quantum limit, $\lambda \to 0$, Eq.~(\ref{eq:relative_tdse}) transforms to Eq.~(\ref{eq:com_tdse}) leading to a two-fold degenerate mode with frequency $\omega_{R}$ \cite{degenerate_modes}.
%
%%-----------------------------------------------------------------------------------------------------------------
%% SECOND FIGURE: External excitation, resonance example

\subsection{Resonance excitation. Absorption spectrum} 
We now verify these results by applying the resonance excitation (II), i.e. using a periodically modulated trapping potential of the form
$\Omega^2[1 + \beta(t)]{\bf{r}}_i^2/2$, with  $\beta(t)=\beta_0 \exp[-(t-t_0)^2/2\sigma] \sin(\omega_{\textup{ext}} t)$ and $i=1,2$. 
Since we expect a strong resonance for 
$\omega_{\textup{ext}}=\omega_{r}$ and $\omega_{\textup{ext}}=\omega_{{R}}$ we use a small modulation depth  and a finite pulse width, typically $\beta_0=5\cdot 10^{-3}$, $t_0=240$ and $\sigma=100$. The system response is  characterized by the behavior of the total energy, $E_{\textup{tot}}$, see
Fig.~\ref{fig:resonance} for typical examples. For a quantitative analysis we use the value  $E^{\infty}_{\textup{tot}}$ sufficiently long after the excitation (cf. the dotted horizontal lines).

\begin {figure}
\includegraphics[width=0.49\textwidth]{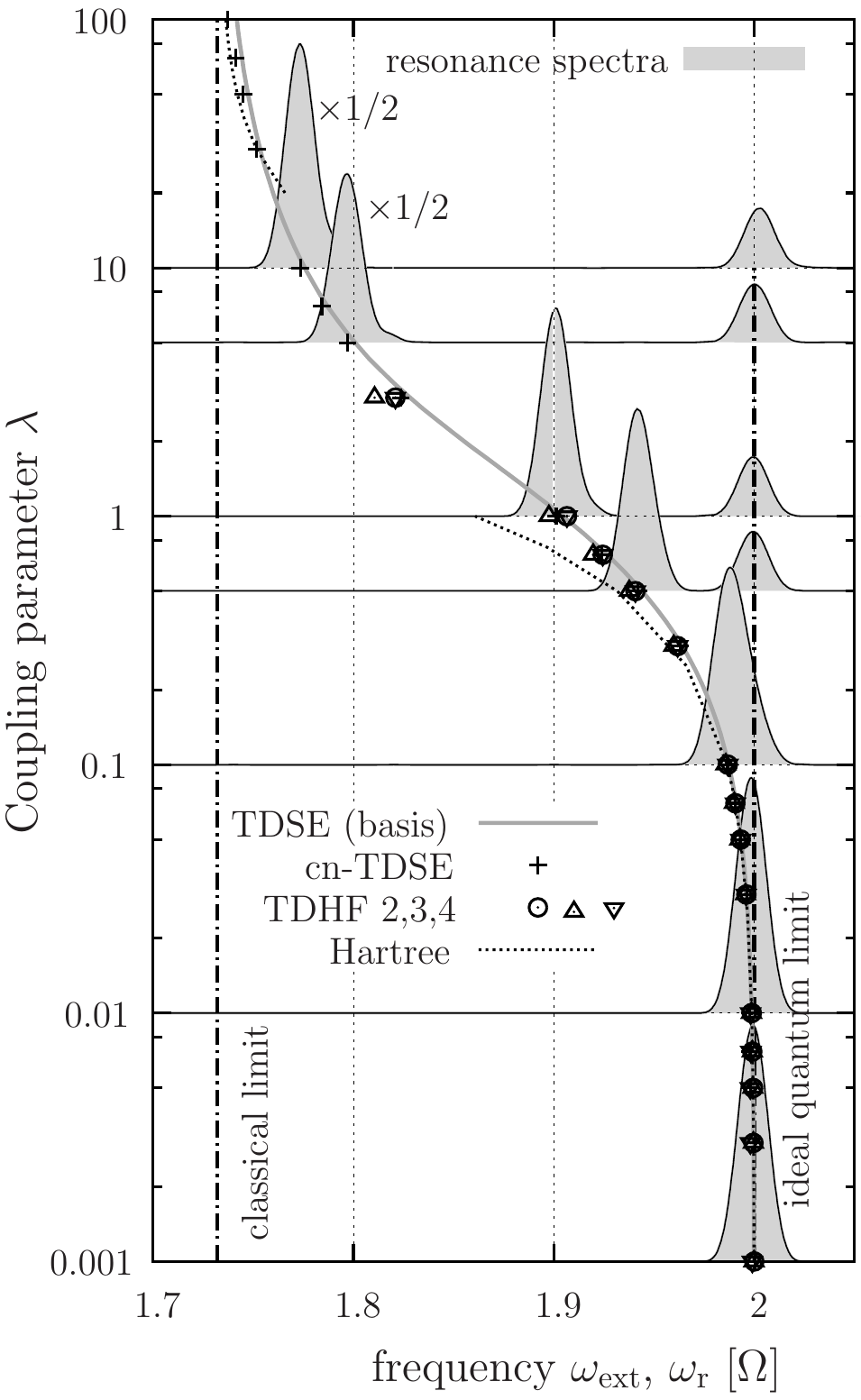}
\caption{QBM frequencies $\omega_r$ and $\omega_R$ for $N=2$ particles in a 1D trap versus $\lambda$ from solution of Eq.~\eqref{eq:relative_tdse} with a basis expansion method (grey line) and a cn-solution of Eq.~\eqref{eq:twopart_tdse}, (+) using excitation method (I). Grey resonance spectra ($E^{\infty}_{\textup{tot}}$) are the result of method (II). Dashed lines: classical ($\lambda > 30$) and quantum  ($\lambda<1$) mean-field models, symbols: TDHF results for $N=2,3,4$, which coincide for $\lambda \leq 1$.}
\label{fig:1d_breathing}
\end {figure}

We now construct the absorption spectra by recording the values $E^{\infty}_{\textup{tot}}(\omega_{\textup{ext}},\lambda)$, by scanning the frequency $\omega_{\textup{ext}}$ for a given $\lambda$ and repeating this procedure for a broad range of $\lambda$ values. The resonance peaks are shown in Fig.~\ref{fig:1d_breathing} and confirm the existence of two frequencies. Their values fully agree with the result of method (I), cf. the grey line.
Moreover, from the peak areas we can deduce the relative spectral weight of the two modes, clearly showing the continuous transition from the ideal quantum limit (where the mode $\omega_{\textup{R}}$ dominates) to the strongly coupled classical case. Both peaks merge at weak coupling, around $\lambda=0.01$ (for the given $\sigma$). On the other hand,  even for $\lambda$ as large as $20$, the pure quantum mode $\omega_{\textup{R}}$ is clearly observable, being of the same absolute importance as for $\lambda=0$.  

%% Neu
Furthermore, we underline that our approach of solving the full two-particle TDSE is applicable beyond the linear response regime. It allows us to estimate the exact energy absorption spectrum including the line shapes which obviously depend on the type of the external pertubation used. This is of special importance for experimental situations, where every excitation has finite duration and typical modulations which will also result in characteristic line shapes.

%
%%-----------------------------------------------------------------------------------------------------------------
%% THIRD FIGURE: 1D Breathing mode (large overview)
%%-----------------------------------------------------------------------------------------------------------------
\subsection{Analytical approximations for $\omega_{r}$}
%In Fig.~\ref{fig:1d_breathing} we also show results from a Hartree model (dashed lines). These perturbation results are applicable in the two limits, respectively, corresponding to strong and weak coupling, and show that the non-perturbative TDSE solution is indispensable for $0.3 \lesssim \lambda \lesssim 30$.

In addition to the numerical results it is possible to derive semi-analytical expressions for the breathing 
frequency $\omega_{r}$ in the two limits of weak and strong coupling, respectively.
To obtain $\omega_{\textup{r}}$ for small $\lambda$, we use a quantum mechanical Hartree model. Since for  $\lambda \ll 1$, the interaction potential $w$ is a small perturbation it can be approximated by the ansatz
\begin {equation}
w({\bf r}_1 - {\bf r}_2) \approx \frac {1}{2} \sum_{i\neq j} \int \textup{d}^3 {\bf r} \; w({\bf r}_i - {\bf r}) |\phi^{\lambda=0}_i({\bf r})|^2 ,
\end {equation}
 involving the densities of the ideal (undisturbed) one-particle states. 
%In the range of non-vanishing density, the sum $(u+v)({\bf r}_1, {\bf r}_2)$ is not necessarily monotonic for $|{\bf r}_i|$ but gives rise to an 
This potential, together with the external potential, gives rise to a renormalized confinement with the
effective harmonic trap frequency $\Omega^ {\textup{eff}}$ yielding the QBM frequency $\omega_{r} = 2\Omega^{\textup{eff}}$. As can be seen in  in Fig.~\ref{fig:1d_breathing} the result is very close to the exact numerical solution of the TDSE for $\lambda \lesssim 0.3$, cf. the dotted line.

Similarly, for large $\lambda$, we construct a semi-classical mean-field theory, where the particles have a finite width and are modelled by Gaussian densities $n(x)$. For any $\lambda$ the corresponding widths are taken from exact diagonalization calculations yielding $\omega_{r}=1+2\lambda V''(|d_0|)$, with the mean-field potential 
\begin {equation}
V(|d|)=\int \; \textup{d}x_1 \textup{d}x_2 \frac{n(x_1) n(x_2)} {x_1-x_2 + |d|},
\end {equation}
and the pertinent equilibrium distance $d_0$. Fig.~\ref{fig:1d_breathing} shows that this approximation works well for $\lambda \gtrsim 30$ (dashed line).

One advantage of these approximations is that they can be straightforwardly extended to larger particle numbers. Yet the most interesting parameter range where quantum and correlation effects are strong simultaneously (around $\lambda=1$) are not accessible by perturbation approximations and require a full numerical treatment.

%%--------------------------------------------------------------------------------------------------------------------------------------------------
%% FOURTH FIGURE: Bosonic wave function, kappa dependence
\begin {figure}
\includegraphics[width=0.49\textwidth]{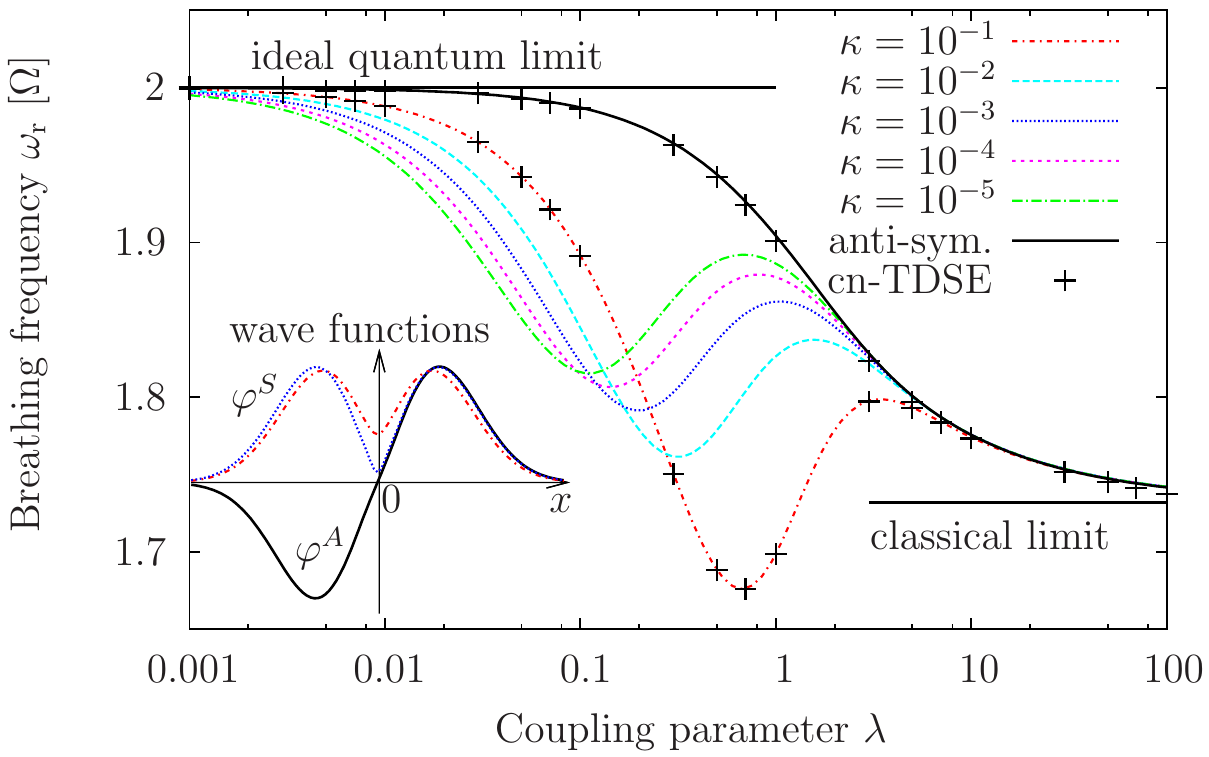}
\caption{(color online) Frequency $\omega_r$ for $2$ particles with a symmetric initial state in a 1D trap from numerical solution of the TDSE using a softened Coulomb potential. Crosses correspond to solution method i) and lines to method ii), see Section \ref{numerics_ss}. For $\kappa=0$, $\omega_r$ is the same for symmetric and anti-symmetric states. The result for the anti-symmetric case is shown by the monotonic full (black) line. A finite $\kappa$ drastically influences $\omega_r$ for the symmetric case which is explained by the behavior of the relative wave function at $x=0$ (see inset).}
\label{fig:1d_boson}
\end {figure}

%%------------------------------------------------------------------------------------------------------------------

\subsection{$\omega_{\textup{r}}$ for more than two particles}
Let us now discuss the dependence of $\omega_{\textup{r}}$ on the number of particles $N$. In the case of classical charged particles in a harmonic trap it is well known that the breathing frequency is universal, i.~e.~$\omega_{\textup{r}}$ does not depend on $N$. This case describes the limit of strong coupling, $\lambda \rightarrow \infty$. On the other hand, for the quantum case at finite $\lambda$, this question is still open. 
It turns out that a numerical solution of the full time-dependent $N$-particle Schr\"odinger equation for $N\ge 3$ with a sufficient accuracy is computationally very expensive. Therefore, we performed instead time-dependent Hartree-Fock calculations for three and four particles using a harmonic oscillator representation of the HF equations \cite{Balzer2009}, see the symbols in Fig.~\ref{fig:1d_breathing}. Full agreement with the TDSE results is observed for $N=2$, up to $\lambda \sim 3$. Further, the data for $N=3,4$ are very close to the TDSE results for $N=2$ up to $\lambda \sim 1$. From this we expect that also the exact $\omega_{\textup{r}}(\lambda)$ is only very weakly $N$-dependent as in the ideal quantum ($\lambda=0$) and classical limits \cite{Henning2008}. However, a conclusive answer to this question requires a thorough computational analysis on the basis of the TDSE for larger particle numbers.
%%-----------------------------------------------------------------------------------------------------------------

%%-----------------------------------------------------------------------------------------------------------------
\section{Influence of the spin statistics on the breathing mode}
\subsection{One-dimensional system}
Let us now repeat the above calculations with excitations (I) and (II), now starting from a symmetric initial coordinate wave function $\Psi_0^{S}$. According to the Bose-Fermi mapping in $1d$ \cite{girardeau} it is expected, that both initial states, $\Psi_0^{S}$ and $\Psi_0^{A}$, should lead to the same QBM. However, it is interesting to note that both numerical approaches to solve the TDSE (see Sec.~\ref{numerics_ss}) fail to reproduce this behavior, due to the singularity of the Coulomb potential. The standard ``regularization'' procedure which introduces a small finite cut-off $\kappa$ in the potential, $\lambda/\sqrt{(x_1-x_2)^2 + \kappa^2}$, yields qualitatively wrong results for $\omega_{r}$, cf. Fig.~\ref{fig:1d_boson}. The breathing 
frequency $\omega_{r}$ exhibits a nonmonotonic dependence on $\lambda$, in contrast to the solution of the 
TDSE for an anti-symmetric initial state. Reducing the cut-off parameter $\kappa$ slightly improves the 
behavior by reducing the amplitude of the oscillation and shifting it towards smaller values of $\lambda$.
But even for $\kappa$ as small as $10^{-5}$ the spurious oscillation persists for $\lambda \lesssim 1$, and no convergence to the behavior of the anti-symmetric state is observed.

We underline that this is not a numerical error and not due to the solution procedure, but it is a property of the regularized Coulomb potential in $1D$. The reason for the observed unexpected behavior is the incorrect (finite) value of the relative wave function $\varphi_0^{S}$ at the origin (i.e. at zero particle separation). This reduces the particle repulsion and, thus, $\omega_{r}$, cf. inset of Fig.~\ref{fig:1d_boson} \cite{as_cn}.

\subsection{2D and 3D systems}
\begin{figure}
 \includegraphics[width=0.48\textwidth]{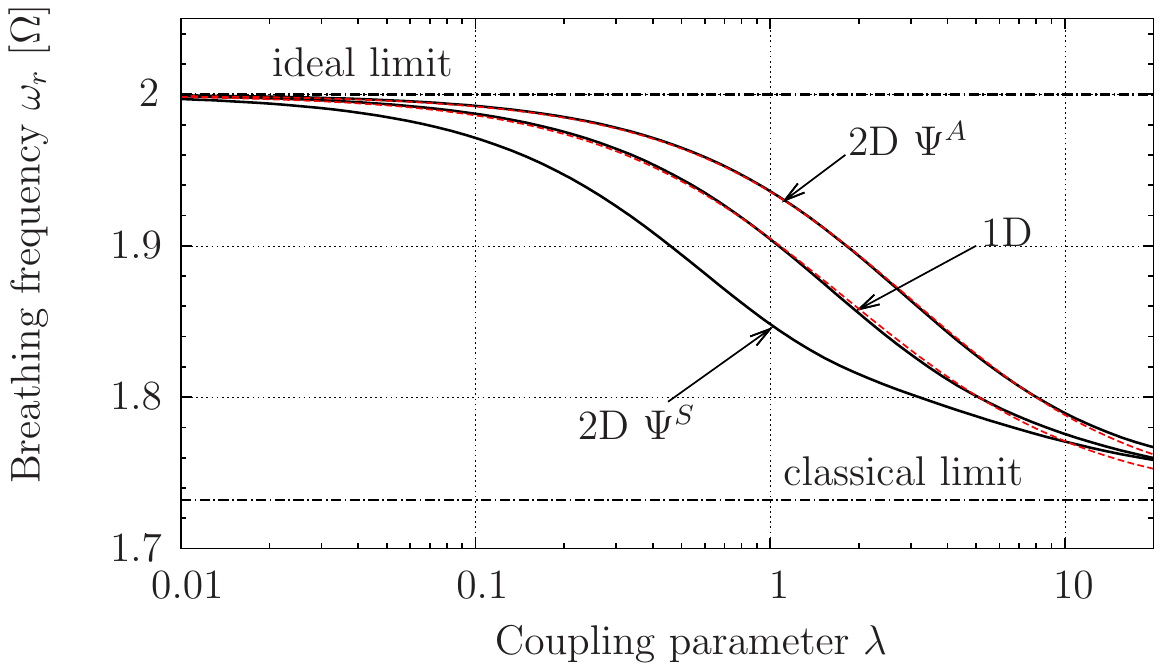}
  \caption {(color online) QBM frequency $\omega_r$, for $N=2$ in a {\em 2D trap} for symmetric and antisymmetric states from solution of the TDSE \eqref{eq:relative_tdse} with $\kappa=0$. For comparison, also the $1D$ result and the two fits $w^{fit}_{\textup{r}}(\lambda)$, Eq.~(\ref{fit}), are shown by the (red) dotted lines.}
 \label{fig:2d_breathing}
\end{figure}

The situation changes completely in $2D$ and $3D$. Now particles can avoid each other, allowing for finite values  $\varphi_0^{S}(0)$, whereas  $\varphi_0^{A}(0)$ remains equal to zero. Consequently, we expect a lowering of $\omega_{r}$ for symmetric states, compared to anti-symmetric ones, for arbitrary $\lambda$. This is fully confirmed by numerical solutions of the TDSE \cite{2d},  cf. Fig.~\ref{fig:2d_breathing}. The overall behavior is the same as in $1D$ --- we observe a QBM with $\omega_{R}=2$ and a second mode $\omega_{r}(\lambda)$, however, 
its value is different from $1D$: the frequency is reduced (increased) for an (anti-)symmetric state,  cf. Fig.~\ref{fig:2d_breathing}.  While the differences vanish in the classical (and quantum) limits due to missing (complete) wave function overlap, at intermediate couplings,  around $\lambda=1$, the differences between $1D$ and $2D$, as well as between anti-symmetric and symmetric states are substantial, reaching values of about $3\%$ and $5\%$, respectively. 
These differences render the QBM frequency $\omega_{r}(\lambda)$ a sensitive diagnostics of the spin statistics of the particles. Indeed, in case of a symmetric spin wave function (e.g. spin-polarized system in a strong magnetic field), the anti-symmetric (symmetric) relative coordinate wave function $\varphi_0$ refers to fermions (bosons) and, vice versa, in case of an anti-symmetric spin wave function. Furthermore, in case of a mixture of fermions and bosons, the resonance absorption  (II) yields information about the fraction of the different components.

 While our results have been obtained for systems with a spin-independent hamiltonian, it is straightforward to 
perform analogous TDSE calculations with spin effects included. 
Another interesting observation is that $\omega_{\textup{r}}(\lambda)$ in $1D$ and $2D$ for the wave function $\varphi_0^A$ is well described by the functional form   
\begin{eqnarray}\label{fit}
\omega^{fit}_{r}(\lambda) &=& a \exp[-\arctan (b \lambda + c)]+d, \\
\nonumber
\mbox{with}\quad  d &=&  (2-\sqrt{3}d_c)/(1- d_c), \\
\nonumber
\quad d_c &=& \exp(\pi/2-\arctan c),\\
a &=& (\sqrt{3} -d) \exp(\pi/2),
\nonumber
\end{eqnarray}
cf. Fig.~\ref{fig:2d_breathing}, \cite{fit}.

Finally, we expect that the reported collective behavior will be observable also in anisotropic harmonic traps in $d$ dimensions with the difference that there will be a total of $2d$ QBMs with frequencies $\omega^i_R$ and $\omega^i_r(\lambda)$ where $i=1, \dots d$.
%%-----------------------------------------------------------------------------------------------------------------
%% FIFTH FIGURE: Two-dimensional breathing mode, boson/fermion difference
%%-----------------------------------------------------------------------------------------------------------------

%% Conclusions
\section{Summary and discussion}
In this paper we have presented a complete analysis of the quantum breathing modes of interacting charged particles in a $1D$ and $2D$ harmonic trap for arbitrary values of the coupling strength $\lambda$. The results are based on accurate solutions of the TDSE with two independent excitation scenarios and differ qualitatively from previous results based on hydrodynamic models. We have shown that the frequency ratio $\gamma(\lambda)=\omega_r(\lambda)/\omega_R$, while being independent of the trap frequency and the confining system, does depend on the trap dimensionality and on fundamental properties of the particles: their spin statistics and the form of their pair interaction. Furthermore the $\lambda$-dependence of $\gamma$ is a sensitive measure of the strength of many-body effects in the particle ensemble.

The present results have been obtained for two particles with Coulomb interaction where exact solutions of the TDSE are possible. Nevertheless, they are representative for any nonideal quantum system in a harmonic trap. An analogous but different $\lambda-$dependence exists for dipole interaction \cite{l23} and for short-range potentials. Thus, for a given value of $\lambda$, $\omega_r$ is indicative of the form of the pair interaction.

While our results are of importance for any nonideal quantum system in a harmonic trap, they are of particular relevance for electrons \cite{afilinov-etal.prl01} and excitons \cite{ludwig_cpp07} in quantum dots as well as for Bose and Fermi systems and their mixtures in traps or optical lattices. There the resonance absorption by the QBM, in particular, the relative spectral weight of the two modes, may serve as a valuable experimental diagnostics. Thereby one can take advantage of the comparatively easy excitation of the breathing mode using the excitation scenarios I or II.

At the same time the presented precise values of $\gamma(\lambda)$ provide a strong benchmark for nonequilibrium theoretical models and computational many-body methods for finite systems, including hydrodynamics, TDHF and time-dependent density functional theory and nonequilibrium Greens functions \cite{Balzer2009,Balzerprb2009}.
Indeed, the normal modes of trapped systems are an important criterion for the quality of nonequilibrium simulations which treat the interaction approximately. This is in complete analogy to approximate simulations of atoms which should reproduce the excitation energy spectrum as accurately as possible. Recently it was shown that the sloshing (Kohn) mode of a harmonically trapped system  is preserved exactly by any conserving many-body approximation \cite{Bonitzprb2007}. Similarly, the breathing mode may serve as such a constraint. For example, as was shown in Fig.~\ref{fig:resonance}, the time-dependent Hartree-Fock approximation behaves very well for an anti-symmetric initial
state (fermions). For symmetric states (bosons) Hartree-Fock simulations of strongly correlated charged systems in a harmonic trap have recently been developed \cite{heimsoth_pe2009} but the results for the breathing mode appear to be significantly less accurate than for fermions.

Finally, a particularly remarkable observation made in our paper is the persistence of the center of mass quantum breathing mode with frequency $\omega_R$ up to very large $\lambda$ values. In strongly coupled systems this mode shows up in a ``breathing'' of {\em each individual particle} (see lower left part of Fig.~\ref{fig:resonance} and the accompanying video \cite{video}). Despite the fact that for strong coupling, the collective quantum degeneracy effects are weak (the interparticle distance is much smaller than $l_0$) each individual particle clearly ``remembers'' its quantum nature.

\begin{acknowledgments}
We acknowledge helpful comments from M. Heimsoth and D. Hochstuhl and financial support by the Innovationsfond  Schleswig-Holstein and Deutsche Forschungsgemeinschaft via SFB-TR 24.
\end{acknowledgments}

\bibliographystyle {apsrev}

\begin{thebibliography}{30}
\bibitem{dipole_comment} We should note, that both excitations, (I) and (II), have vanishing dipole moments. Therefore, no dipole oscillation, as e.g. the well-known Kohn mode, can be excited.
\bibitem{excitation} Note that for other types of excitation other center of mass motions will be observed. In particular, a dipole excitation will give 
rise to the familiar sloshing or Kohn mode which is independent of $N$, dimensionality, interaction or quantum and spin effects.
\bibitem{degenerate_modes} For $1/r^2$ interaction the existence of a two-fold degenerate BM with $\omega_R=\omega_r=2$ was reported in Ref. \cite{geller96}.
\bibitem{as_cn} In contrast,  $\varphi_0^A$ vanishes at the singularity, and the numerical result for $\omega_{r}$ is correct, in the limit of small $\kappa$. Hence, we used $\varphi_0^A$ in our 1D calculations.
\bibitem{2d} In $2D$, the Coulomb integrals converge and no cut-off parameter $\kappa$ is required.
\bibitem{fit} The limits, $\omega_{\textup{r}}(0)=2.0$ and $\omega_{\textup{r}}(\lambda \rightarrow \infty)=\sqrt{3}$ are built-in. In 1D (2D) the parameters are $b=0.867 \;(b=0.204)$ and $c=1.742 \; (c=0.296)$.
\end{thebibliography}

\end{document}